\documentclass{article}

\usepackage{graphicx}

\def\abstract#1{\vskip 7mm
        \begin{center}{\large Abstract}\par \smallskip
                \begin{minipage}[c]{12cm}
                        \small #1
                \end{minipage}
        \end{center}
}
\def\title#1{\begin{center}{\Large\bf #1}\end{center}}
\def\author#1{\vskip 5mm \begin{center}{#1}\end{center}}
\def\address#1{\begin{center}{\it #1}\end{center}}
\makeatletter
\@ifundefined{lesssim}{}{}
\@ifundefined{gtrsim}{}{}
\def\vereq#1#2{\lower3pt\vbox{\baselineskip1.5pt \lineskip1.5pt
\ialign{$\m@th#1\hfill##\hfil$\crcr#2\crcr\sim\crcr}}}
\makeatother

\begin{document}

\title{%
Recent progress in wormhole dynamics }
\author{%
  Sean A. Hayward\footnote{E-mail: hayward@mm.ewha.ac.kr}
}
\address{%
  Department of Science Education, Ewha Womans University, \\
  Seoul 120-750, Korea
}

\abstract{Space-time wormholes were introduced in Wheeler's idea of space-time
foam. Traversible wormholes as defined by Morris \& Thorne became popular as
potential short cuts across the universe and even time machines. More recently,
the author proposed a general theory of wormhole dynamics, unified with
black-hole dynamics. This article gives a brief review of the above ideas and
summarizes progress on wormhole dynamics in the last year. Firstly, a numerical
study of dynamical perturbations of the first Morris-Thorne wormhole showed it
to be unstable, either collapsing to a black hole or exploding to an
inflationary universe. This provides a mechanism for inflating a wormhole from
space-time foam to usable size. Intriguing critical behaviour was also
discovered. Secondly, a wormhole solution supported by pure radiation was
discovered and used to find analytic examples of dynamic wormhole processes
which were also recently found in a two-dimensional dilaton gravity model: the
construction of a traversible wormhole from a Schwarzschild black hole and vice
versa, and the enlargement or reduction of the wormhole.}

\section{Introduction}

Space-time wormholes were first described as such by Wheeler \cite{W}, who
envisaged the still-popular space-time foam: the smooth space-time of General
Relativity suffering quantum-gravitational fluctuations in topology at the
Planck scale, becoming a continual foam of transient interconnections.
Curiously, all the standard black-hole solutions have a wormhole spatial
topology, with a minimal surface connecting two asymptotically flat regions,
our universe and a mirrored universe. For the simplest Schwarzschild black
hole, this Einstein-Rosen bridge was properly understood only much later, also
by Wheeler, as a non-traversible wormhole: the two universes are not in causal
contact, with any attempted crossing leading only into the black hole. (See
last year's proceedings \cite{j11} for space-time diagrams and explanations
mostly not repeated here). Later still, Morris \& Thorne \cite{MT} proposed
traversible wormholes, which have similar spatial geometry, with a minimal
surface connecting two asymptotically flat regions, but such that the minimal
surface (the wormhole throat) is preserved in time. Thus a traveller can cross
between the two universes at will. Such wormholes became popular both in
science fiction and as a research topic \cite{V}, apparently also allowing
time-machine construction. However, most research has concerned static or
cut-and-paste wormholes, or lacked an independently defined exotic matter
model. In contrast, the dynamical behaviour of wormholes has received little
attention until recently, and it is only in the last year or two that exact
solutions describing wormhole construction and enlargement have been found.

\section{Static and dynamic wormholes}

Morris \& Thorne's unusually accessible article gave a list of criteria for
traversible wormholes which is still worth consulting. Paraphrasing briefly,
the basic wormhole criteria were: (1) a static, spherically symmetric
space-time; (2) the Einstein equation; (3) a throat (minimal surface)
connecting two asymptotically flat regions; and (4) no (Killing) horizon, so
that the wormhole allows two-way travel. The usability criteria were (5) small
tidal gravitational forces and (6) small proper time for crossing, by human
scales. Finally, the criteria which were mentioned but largely left open were:
(7) physically reasonable matter; (8) perturbative stability, e.g.\ due to a
spaceship; and (9) that it should be possible to assemble the wormhole. These
are, of course, the more interesting issues.

Recent work can be summarized by a corresponding list. (1) One needs to
generalize to any space-time, without symmetry, in particular to non-static
space-times. (2) Although Einstein gravity will be assumed in this article,
many alternative gravitational theories allow traversible wormhole solutions,
including scalar-tensor theories and brane-world models. The most fundamental
point is that one needs to (3) generalize and localize the definition of the
throat and (4) impose local two-way traversibility. Such a definition was given
by the author \cite{wh} in terms of trapping horizons, which are hypersurfaces
foliated by marginal surfaces, which are extremal surfaces in a null
(light-like) hypersurface \cite{bhd}--\cite{mg9}. They can be locally
classified as future or past and outer or inner. Examples include the outer and
possible inner horizons of black and white holes. For a Morris-Thorne wormhole,
the throat is a double outer trapping horizon; double since the minimal
surfaces are extremal in both null directions, and outer since the extremal
surface should be minimal rather then maximal, encoding the so-called flare-out
condition. The doubled nature of the throat indicates that a non-static
wormhole will generally have two mouths, defined by outer trapping horizons,
connected by a tunnel of trapped surfaces. In the static case, the tunnel
shrinks away and the mouths coincide as the throat. A similar understanding was
reached contemporaneously by Hochberg \& Visser \cite{HV}, though their
flare-out condition can generically select maximal rather than minimal surfaces
in a time-symmetric hypersurface.

The author proposed that a black hole can be locally characterized by an
achronal (spatial or null) future outer trapping horizon, and a traversible
wormhole by two temporal (time-like) outer trapping horizons in mutual causal
contact \cite{wh}. Note that the main difference is the causal nature of the
horizon, which is respectively one-way or two-way traversible, as expected in
each case. The Einstein equation then shows that they occur respectively under
positive and negative energy density, specifically referring to the null energy
condition \cite{wh}. This means that they are supported respectively by normal
matter or vacuum, and what was dubbed exotic matter \cite{MT}. This explains
why black holes are common astrophysically, while wormholes have not been
observed. However, the recently discovered acceleration of the universe implies
that cosmic evolution is dominated by unknown dark energy, which violates at
least the strong energy condition. Also, it is well known that negative energy
densities are endemic in quantum field theory. Whatever the identity of the
exotic matter, there is a unified framework for traversible wormholes, black
holes and white holes \cite{wh}--\cite{mg9}. Since the causal nature of the
horizons depends on the sign of the energy density, which may change with time,
black holes and wormholes should theoretically be interconvertible. A famous
example is a Schwarzschild black hole evaporating by Hawking radiation, which
semi-classically is a traversible wormhole by any reasonable definition
\cite{j11,wh}.

Returning to the Morris-Thorne criteria, the usability criteria (5)--(6) are
already sufficiently general. However, any concrete study of wormhole dynamics
requires (7) a specific exotic matter model, so that there are field equations
to determine the evolution. Most early studies ignored the identity of the
exotic matter \cite{MT,V}, but authors increasingly address this issue. Much
work has concentrated on alternative gravitational theories such as
scalar-tensor theories, where static wormhole solutions are found to be common.
Other work draws the exotic matter from quantum field theory in a
semi-classical approximation. The author has proposed using simple exotic
matter models, in order to study issues of principle which were hardly
understood. This has led to concrete examples which show conclusively that (8)
wormhole stability depends on the exotic matter model. Although there have been
various studies of linearization stability, dynamic stability has been
investigated in only two cases: the HKL wormhole was found analytically to be
dynamically stable \cite{HKL}, while the first Morris-Thorne wormhole was found
numerically to be dynamically unstable \cite{SH}. Finally, concerning (9) how
to assemble or construct a wormhole: irradiating a CGHS black hole with
negative energy converts it to an HKL wormhole \cite{HKL,KHK1}, and similarly a
wormhole discovered recently by the author \cite{pur} can be constructed from a
Schwarzschild black hole \cite{KHK2}. The main results of these recent studies
are summarized in the following.

\section{Analytical results}

The first concrete results were found using an exactly soluble model, CGHS
two-dimensional dilaton gravity, generalized to include a massless ghost
(negative-energy) Klein-Gordon field, which supports the existence of static
wormhole solutions \cite{HKL}. One can then set initial data corresponding to
dynamical perturbations of a CGHS black hole or an HKL wormhole, then
analytically find the evolved space-time. As described in last year's
proceedings \cite{j11}, solutions were found describing wormhole collapse to a
black hole, wormhole construction from a black hole, wormhole operation for
transport, including the back-reaction of the transported matter on the
wormhole, and wormhole maintenance, i.e.\ maintaining a static state under
transportation \cite{HKL}.

\begin{figure}
\hskip-5mm
\includegraphics[height=44mm]{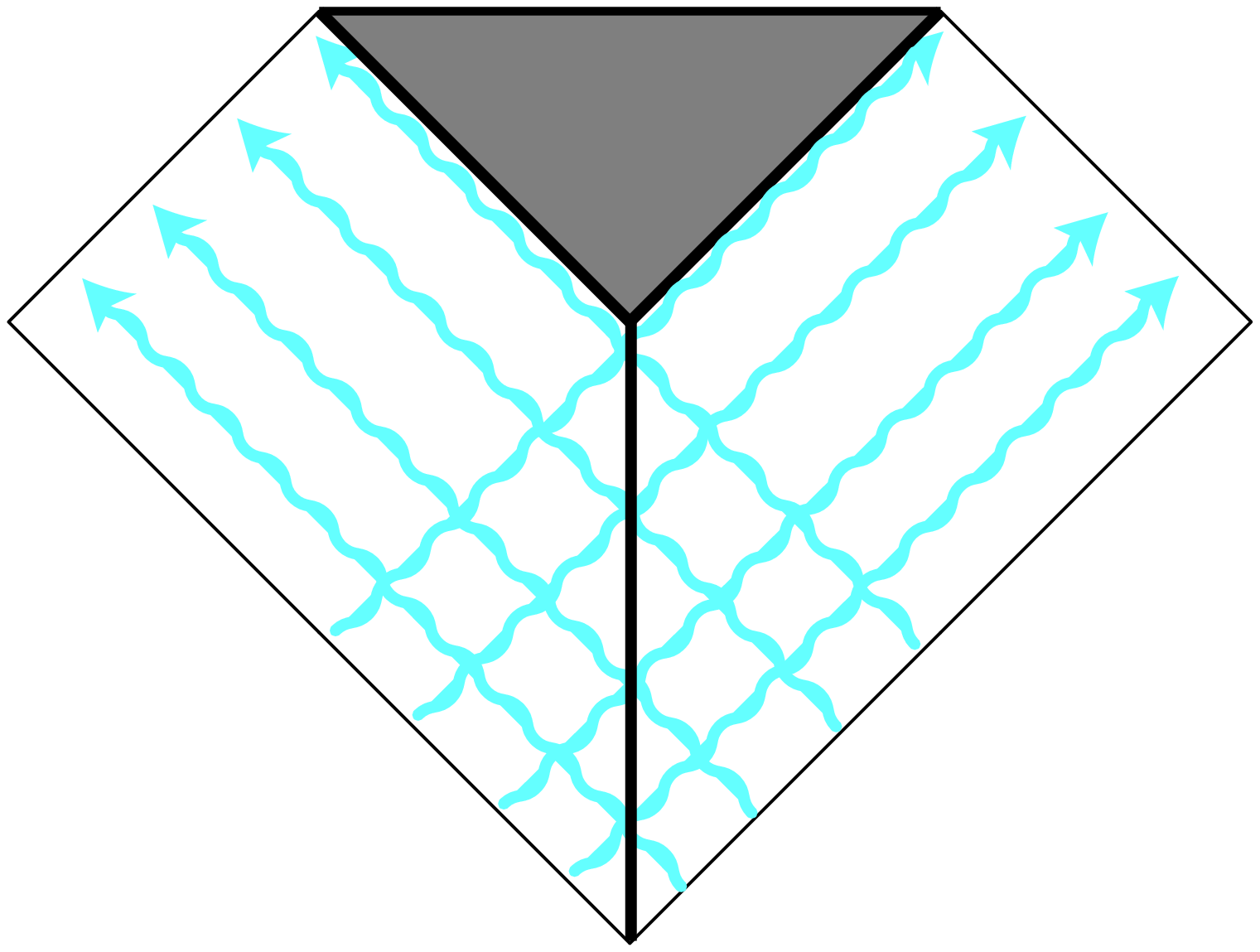}
\includegraphics[height=44mm]{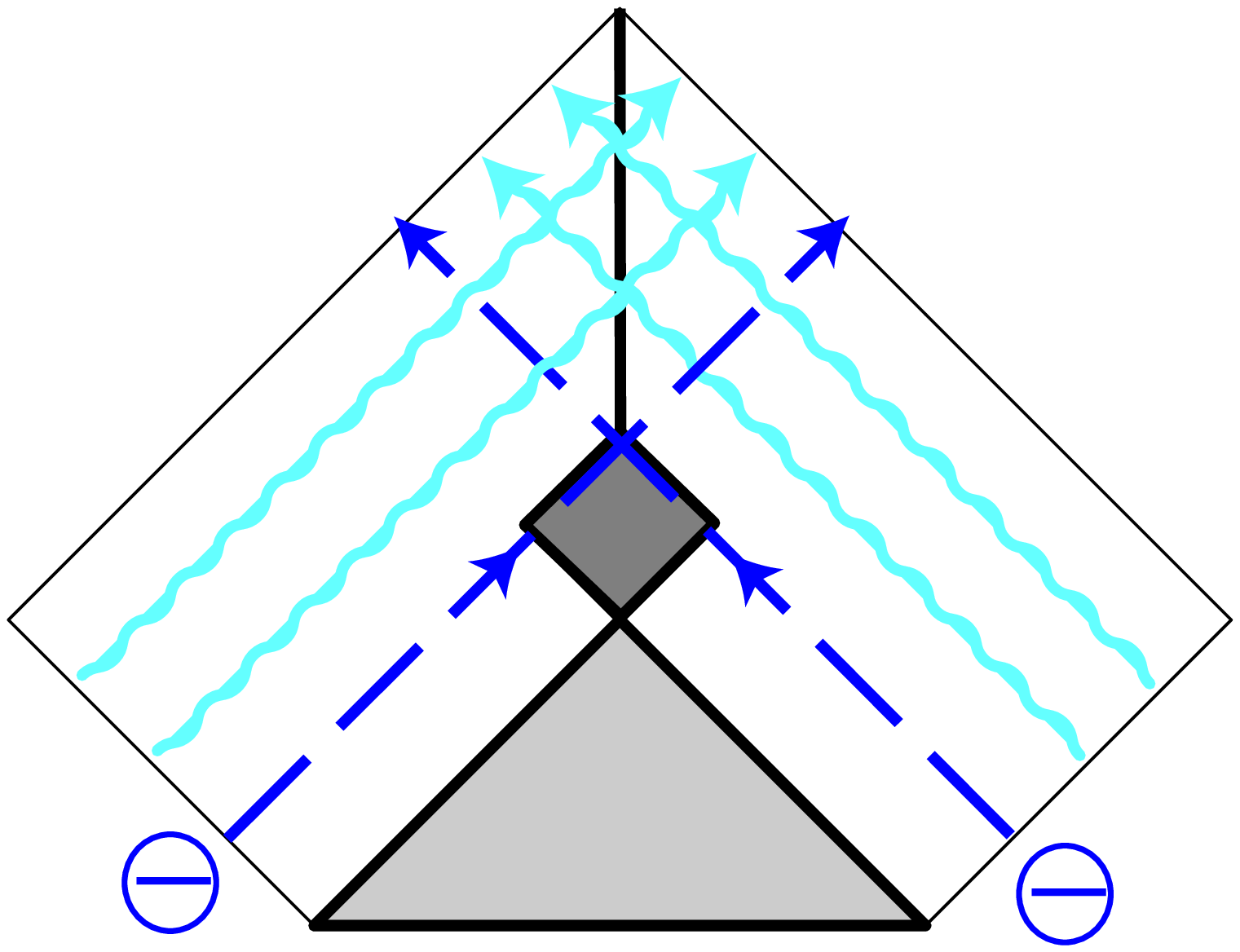}
\includegraphics[height=44mm]{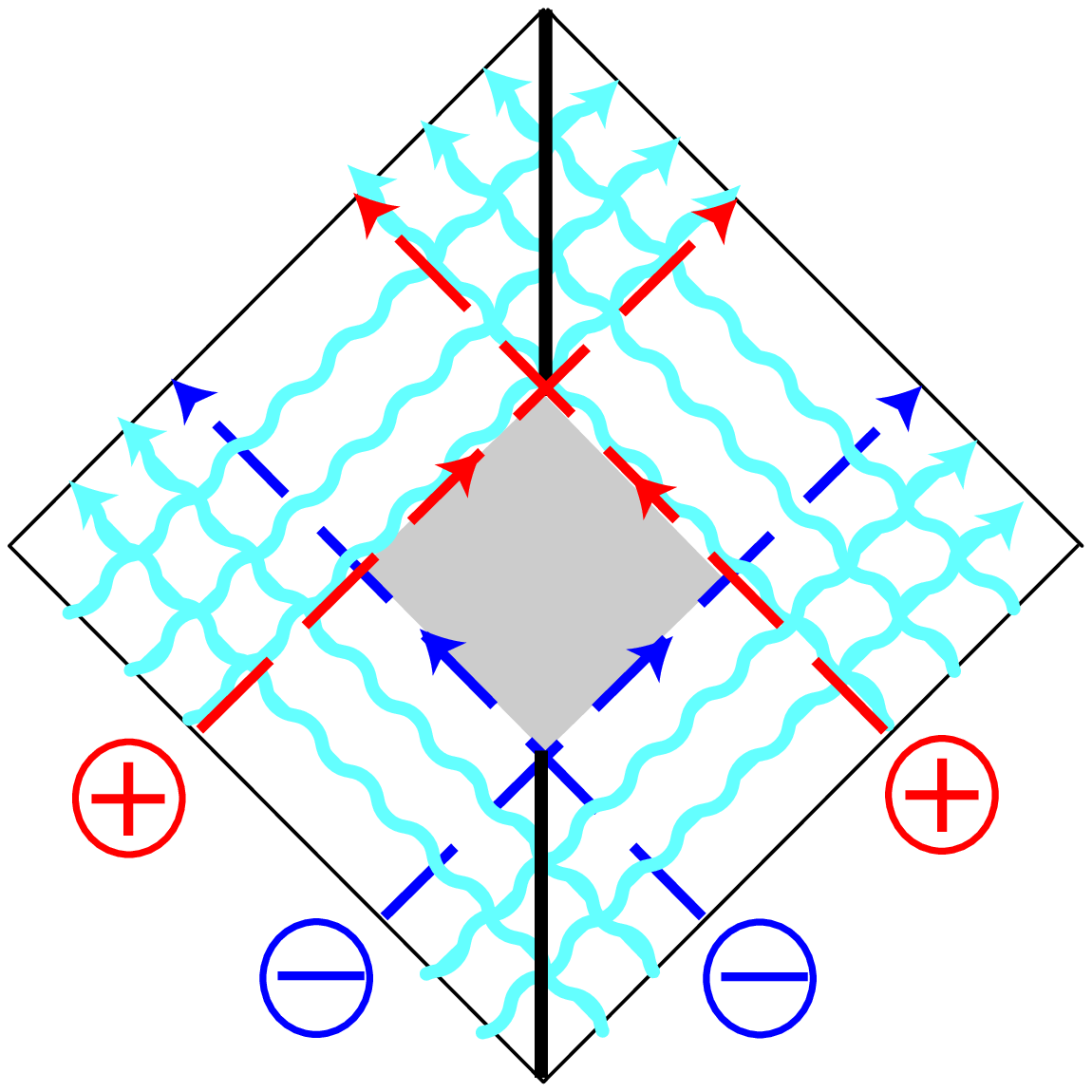}
 \vspace{-5mm}
 \caption{Penrose space-time diagrams. (i) A wormhole collapses to a black hole
 if the supporting ghost radiation, represented by wavy lines, is switched off.
 (ii) A wormhole is constructed from a black hole by irradiating with impulsive
 ghost radiation, represented by dashed lines, followed by constant
 non-impulsive radiation to support the resulting wormhole. (iii) A wormhole is
 enlarged by beaming in impulsive ghost radiation, switching off the
 non-impulsive radiation, then beaming in compensating impulsive normal
 radiation. The bold lines represent the trapping horizons, light shading
 indicates past trapped regions and darker shading indicates future trapped
 regions. The impulses discontinuously shift the trapping horizons constituting
 the black-hole horizons or the wormhole throat. In the dilaton gravity model,
 this refers to the CGHS black hole and the HKL wormhole, and in full Einstein
 gravity to the Schwarzschild black hole and a recently discovered wormhole.}
 \label{fig1}
\end{figure}

In order to study similar dynamic processes in full Einstein gravity, the
author proposed a simple exotic matter model, pure ghost radiation, i.e.\ pure
radiation with negative energy density \cite{j11}. Wormhole solutions of the
Morris-Thorne type were found \cite{pur}, supported by equal left-moving and
right-moving radiation. Since the radiation propagates without interaction or
backscattering, it is easy to see what happens if it is suddenly switched off
from both sides of the wormhole: the wormhole collapses to a black hole, as in
Fig.\ref{fig1}(i). The radiation escapes in Vaidya regions, leaving a vacuum
Schwarzschild region, with continuity determining that the wormhole throat
bifurcates to form the black-hole horizons.

For other dynamic processes, a useful idealization is impulsive radiation,
where the radiation is concentrated so as to deliver finite energy and momentum
in an instant. In the two-dimensional model, further examples were given of
wormhole construction and operation, plus how to stably enlarge a wormhole
\cite{KHK1}. Corresponding versions of some of these processes can be found
analytically in full Einstein gravity by matching Schwarzschild, Vaidya and
static-wormhole regions \cite{KHK2}. In particular, irradiating a Schwarzschild
black hole with impulsive then constant ghost radiation can convert it to a
wormhole of the above type, as in Fig.\ref{fig1}(ii). Also, the wormhole can be
enlarged by adding additional impulsive ghost radiation followed by balancing
normal impulsive radiation, as in Fig.\ref{fig1}(iii). The ghost radiation is
switched off between the impulses, so that the middle region is found
analytically as part of a Schwarzschild white hole, therefore expanding and
increasing the wormhole size. The opposite ordering of the impulses would
reduce the wormhole size. These will be the first analytic solutions describing
wormhole construction or enlargement in full Einstein gravity.

\section{Numerical results}

The first numerical study of wormhole dynamics was also performed recently
\cite{SH}. The starting point was a static wormhole which is best known as
Morris \& Thorne's opening example, though it is actually a solution for a
massless ghost Klein-Gordon field, as was shown earlier by various authors. The
idea was to study dynamical perturbations of this static wormhole, using the
spherically symmetric Einstein system with the above exotic matter model, and
also a normal massless Klein-Gordon field, to see the effect on the wormhole of
normal matter, like an astronaut or spaceship traversing the wormhole.

A numerical code was developed based on a dual-null coordinate system, in order
to follow the horizon dynamics and radiation propagation accurately. After
testing the code by evolving the static wormhole, the numerical experiments
added or subtracted Gaussian pulses in the ghost field or the normal field,
parametrized by amplitude, width and position. The space-time evolution was
followed by tracking the trapping horizons and quantities such as area and
energy. The local gravitational energy \cite{1st} is a key indicator; for the
static wormhole it is positive, maximal at the throat and tends to zero at
infinity, so that the total (Bondi or ADM) mass of the static wormhole
vanishes. For dynamic perturbations, the sign and size of the initial Bondi
energy $E_0$ was found to be the main factor affecting the outcome.

\begin{figure}
\includegraphics[height=57mm]{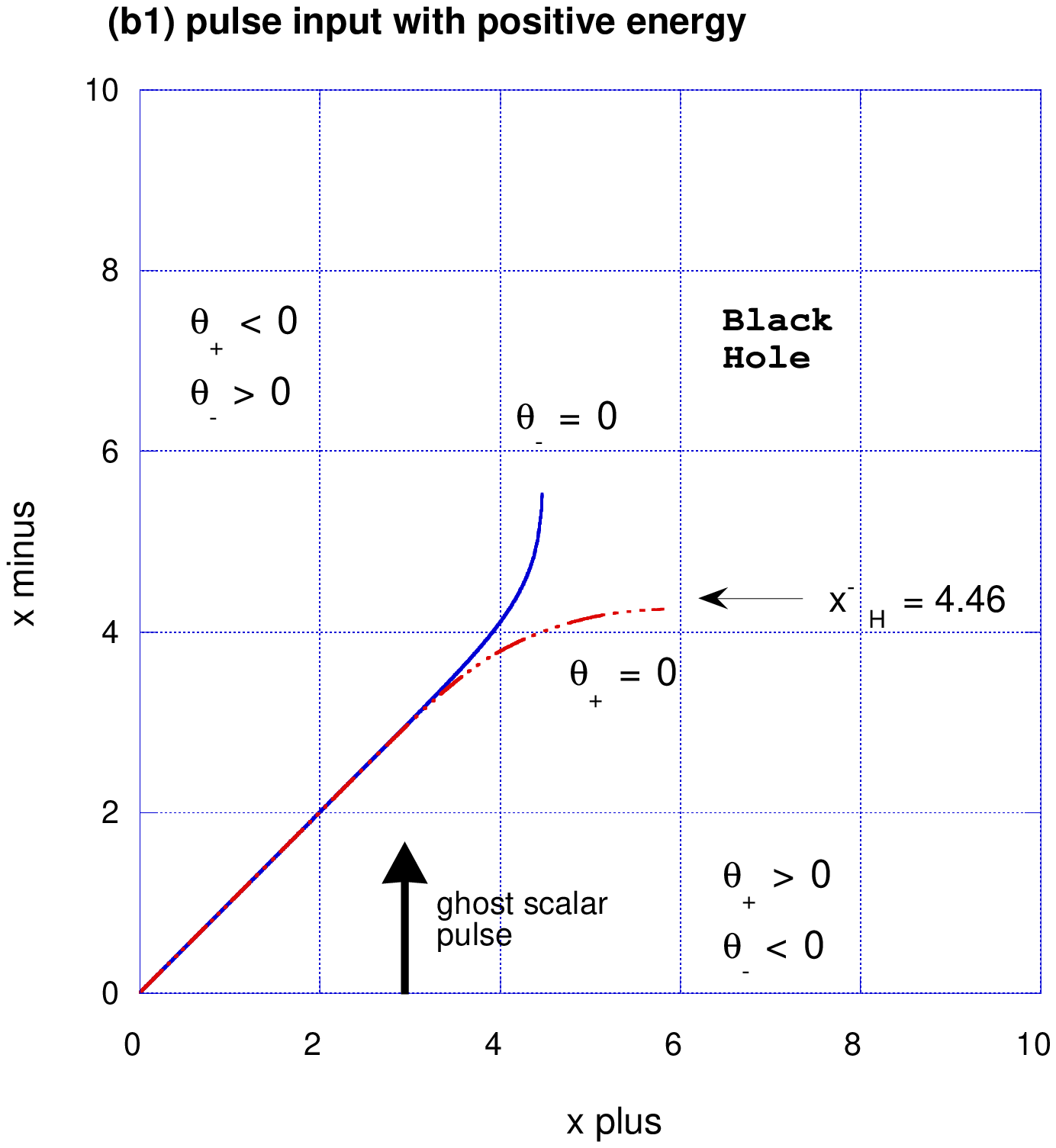}
\includegraphics[height=57mm]{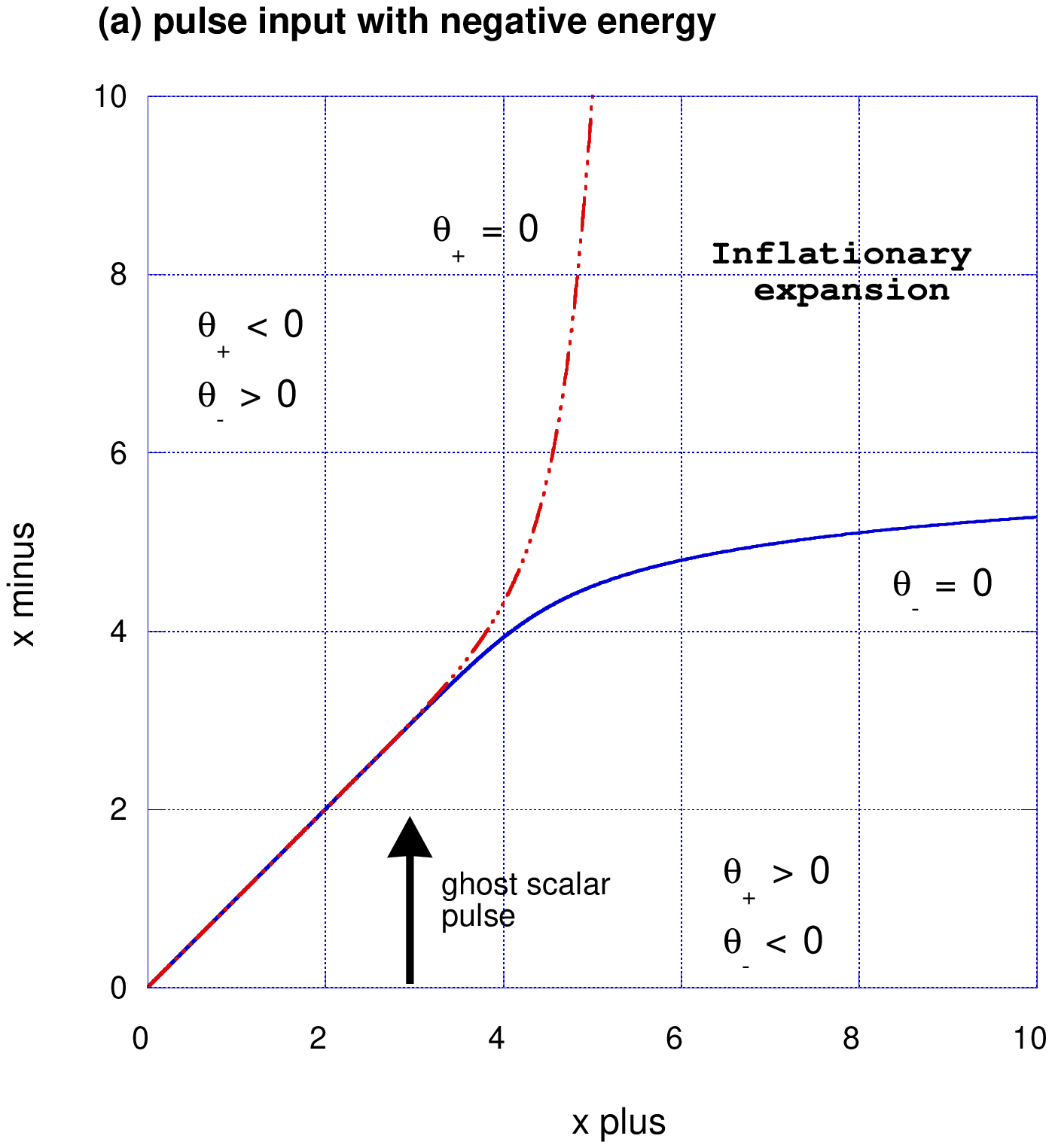}
\includegraphics[height=4cm]{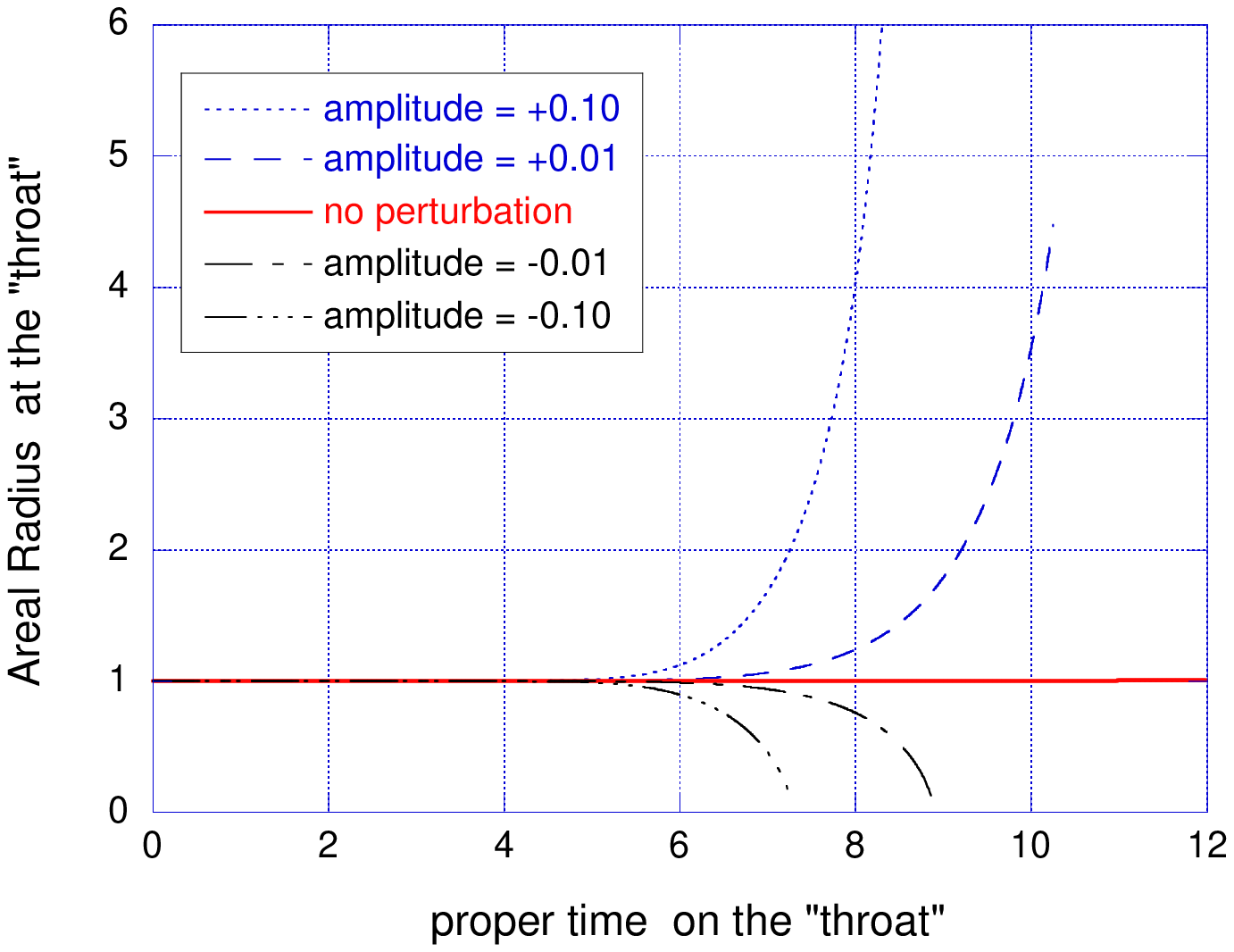}
 \vspace{-5mm}
 \caption{Location of the trapping horizons $\theta_\pm=0$ when the static
 wormhole is perturbed by Gaussian pulses with (i) positive energy and (ii)
 negative energy. The axes are the dual-null coordinates $x^\pm$, so that a
 $45^\circ$ counterclockwise rotation of the figure is a partial Penrose
 diagram. (iii) Areal radius of the throat $x^+=x^-$, plotted as a function of
 proper time. Additional negative energy causes inflationary expansion, while
 reduced negative energy causes collapse to a black hole and central
 singularity. For smaller initial energy, the collapse or explosion occurs more
 slowly.}
 \label{fig2}
\end{figure}

Fig.\ref{fig2}(i) shows what happens for a positive-energy pulse, $E_0>0$: when
the pulse hits the wormhole throat, the double trapping horizon bifurcates to
form a tunnel of future trapped surfaces, with the two mouths accelerating away
from each other and approaching the speed of light, forming the horizons of a
black hole. The final mass $M$ of the black hole can be easily extracted due to
rapid convergence in the dual-null system. In summary, the wormhole has
collapsed to a black hole.

Fig.\ref{fig2}(ii) shows what happens for a negative-energy pulse, $E_0<0$:
when the pulse hits the wormhole throat, the double trapping horizon bifurcates
to form a tunnel of past trapped surfaces, with the two mouths accelerating
away from each other and approaching the speed of light, forming the
cosmological horizons of an inflationary universe. The Hubble constant $H$ can
be found by fitting the exponential curve of area against proper time,
Fig.\ref{fig2}(iii). In summary, the wormhole has exploded into an inflationary
universe. This provides the first concrete mechanism for inflating wormholes
from space-time foam to macroscopic size.

For Gaussian pulses in the normal field, the results are similar, except that
the total energy $E_0$ is necessarily positive and so the wormhole always
collapses to a black hole. Thus a traveller successfully crossing the wormhole
would nevertheless look back to discover that the passage has caused the
wormhole to collapse, thereby ironically sealing off the causal connection to
the home universe.

By varying mainly the amplitude of the pulse, perturbations as small as
$E_0\approx\pm10^{-4}a/2$ can be reliably evolved, where $a$ is the initial
throat radius. In all cases, positive or negative initial energy respectively
causes collapse or explosion. Clearly the wormhole is dynamically unstable,
though it had been previously found to be linearization stable, indicating a
non-linear instability. For smaller perturbations, the collapse or explosion
occurs more slowly and the final mass or Hubble constant is smaller. Unexpected
critical behaviour was also discovered: the initial energy determines the
collapse time $x-x_0=-0.60\ln E_0$ and there appears to be a minimal black-hole
mass $M=0.30a$ or Hubble constant $H=1.1/a$ as perturbations tend to zero, for
both exotic and normal field perturbations. All this is unexplained.

\section{Conclusions}
The largely new area of wormhole dynamics has been substantially developed
recently, based on a local, dynamical theory of traversible wormholes
\cite{wh}, with mouths defined by temporal outer trapping horizons, unified
with a local, dynamical theory of black holes \cite{bhd}--\cite{mg9}. Concrete
examples \cite{HKL}--\cite{KHK2} have supported the following conclusions.

\noindent$\bullet$ Traversible wormholes can be constructed from black holes by
absorbing exotic matter.

\noindent$\bullet$ Traversible wormholes can collapse to black holes, by losing
exotic matter or gaining normal matter, such as a traveller or spaceship.

\noindent$\bullet$ Traversible wormholes can explode to inflationary universes,
by gaining exotic matter. This provides a mechanism for inflating wormholes
from space-time foam to usable size.

\noindent$\bullet$ An exotic matter model must be specified for a given
problem. Apart from semi-classical quantum field theory, dark energy models and
alternative gravitational theories, simple models have theoretical merit in
understanding basic principles.

\noindent$\bullet$ Traversible wormholes can be dynamically stable or unstable,
depending on the exotic matter model. Linearization stability is not
conclusive.

\noindent$\bullet$ Stable wormholes can be operated and maintained by balance
of positive and negative energy, and enlarged or reduced by ordering of
positive and negative energy.

\noindent$\bullet$ Numerically discovered critical behaviour suggests critical
black-hole and inflationary-universe solutions.

\noindent$\bullet$ Wormhole dynamics with a specific exotic matter model need
not be plagued by naked singularities or causal loops. Time-machine
construction \cite{MT,V}, often described as ridiculously easy from
(cut-and-paste) wormholes, is still an open question if given field equations
are to be satisfied everywhere. Cosmic Censorship and Chronology Protection may
yet survive physically reasonable exotic matter.

\bigskip\noindent Research partially supported by Korea Research Foundation grant
KRF-2001-015-DP0095. Thanks to the conference organizers for support and Hiroko
Koyama and Hisa-aki Shinkai for supplying figures.


\begin{thebibliography}{99}
\bibitem{}Many important references have been omitted here due to size and
format restrictions.
\end{thebibliography}
\end{document}